\documentstyle[12pt]{article}
\newcommand{\lsim}{\stackrel{\mbox{\raisebox{-0.1ex}{\scriptsize $<$}}}
{\mbox{\raisebox{-0.5ex}{\scriptsize $\sim$}}}}

\newlength{\myleftmargin}
\newlength{\paperwidth}
\def\Journal#1#2#3#4{{#1}{\bf #2}, #3 (#4)}

\def\NPB{{\it Nucl. Phys.} {\bf B}}
\def\PLB{{\it Phys. Lett.} {\bf B}}

\def\PRD{{\it Phys. Rev.} {\bf D}}
\def\ZPC{{\it Z. Phys.} {\bf C}}

\def\etal{{\it et al.}}
\def\ie{{\it i.e.}}

\renewenvironment{thebibliography}[1]
 {\section*{References}
  \begin{list}{\arabic{enumi}.}
    {\usecounter{enumi} \setlength{\parsep}{0pt} \setlength{\itemsep}{2pt}
     \settowidth{\labelwidth}{#1.}
     \setlength{\leftmargin}{\parindent}
     \setlength{\labelsep}{\leftmargin}
     \addtolength{\labelsep}{-\labelwidth}
     \sloppy
    }}{\end{list}}

\begin{document}

\baselineskip=16pt

\renewcommand{\thefootnote}{\fnsymbol{footnote}}
\begin{flushright}
KOBE--FHD--95--03\\
April~~5~~1995
\end{flushright}
\begin{center}
{\bf TWO--SPIN ASYMMETRY OF HEAVY QUARKONIA\\
PRODUCTIONS AT RHIC ENERGY}\\
\vspace{2em}
T. YAMANISHI\footnote[2]{Presented in the 1995 Lake Louise Winter Institute
at Lake Louise, Canada on 19--25 February.
\\E--mail~~~~yamanisi@natura.h.kobe--u.ac.jp}\\
{\it Graduate School of Science and Technology,}\\
{\it Kobe University, Kobe, Japan 657}\\
\vspace{1em}
and\\
\vspace{1em}
T. MORII\footnote[8]{E--mail~~~~morii@kobe--u.ac.jp}, S. TANAKA\\
{\it Faculty of Human Development, Division of}\\
{\it Sciences for Natural Environment,}\\
{\it Kobe University, Kobe, Japan 657}\\
\vspace{2.5em}
ABSTRACT\\
\vspace{0.4cm}
{\footnotesize
\begin{minipage}{30pc}
Hadroproductions of charmonium and bottomonium $\chi_J$ in polarized $pp$
collisions at RHIC energies are studied.
Two--spin asymmetries $A_{LL}^{\chi_J}(pp)$ for these reactions are very good
parameters for examining the polarized gluon distributions in a proton.
\end{minipage}
}
\end{center}
\vspace{0.4cm}
\parindent 3pc

\noindent
{\bf 1~~~Introduction}
\vspace{0.4cm}

Relativistic Heavy Ion Collider (RHIC)\cite{ref:RHIC} which is designed to
have a beam polarization of about 70 \% with a luminosity of 2$\times$10$^{32}$
cm$^{-2}$sec$^{-1}$ at an energy of $\sqrt s=50-500$ GeV
will give us fruitful informations on the spin
structure of proton.
Recent measurements of the spin--dependent proton structure
function $g_1^p(x, Q^2)$ by EMC\cite{ref:EMC88} and SMC
Collaboration\cite{ref:SMC94}, which are far from the Ellis--Jaffe
prediction\cite{ref:Ellis74}, confirmed again the suggestion that very little
of the proton spin is carried by quarks.
Many ideas have been proposed so far to solve this problem.
One of the interesting approach is that the gluon plays a crucial role on
the nucleon spin structure\cite{ref:anomaly}.
But the magnitude $\Delta G$ and
distribution $\delta G(x)$ of the gluon polarization inside a proton is not
known yet. In order to improve our understanding of the nucleon structure,
it is necessary to measure $\Delta G$ and $\delta G(x)$ in other reactions
which are sensitive to the gluon polarization.

Among various reactions, the $\chi_J$--productions in hadron--hadron collisions
are interesting candidates. In this paper, we are concerned with the
$\chi_J$--hadroproductions at low--$p_T$ regions.
The cross sections of these processes are expected to depend sensetively on
the shape of polarized gluon distributions because the gluon--gluon fusion is
the dominant mechanism at the lowest order of QCD (see Figure 1).
Although these productions have been carefully studied so far
by several people in order to extract unpolarized gluon distributions
$G(x)$ and test the perturbative QCD\cite{ref:Baier83}, we are interested in
the behavior of the polarzed gluon distributions $\delta G(x)$.

So far, various types of the polarized gluon distribution functions
$\delta G(x)$ have been proposed: some of them have large $\Delta G$
($\simeq 5 - 6$)\cite{ref:Altarelli89,ref:Cheng90,ref:Rams91,ref:Morii94}
and others have small $\Delta G$
($\lsim 2 - 3$)\cite{ref:Cheng90,ref:Rams91,ref:Brod94,ref:Gehrmann}, where
$\Delta G=\int^1_0 \delta G(x)dx$.
Here we take the following typical types of $x\delta G(x)$ to examine the
effect on the $\chi_J$--hadroproductions:
\begin{enumerate}
\renewcommand{\labelenumi}{(\alph{enumi})}
\item MTY model \cite{ref:Morii94}~:
\begin{eqnarray}
&&x\delta G(x, Q^2=10{\rm GeV}^2)=9.52x^{0.4}(1-x)^{17}~,
\label{eqn:typeA}\\
&&{\rm with}~~\Delta G(Q^2_{SMC})=5.32~,\nonumber
\end{eqnarray}
\item Cheng--Lai model \cite{ref:Cheng90}~:
\begin{eqnarray}
&&x\delta G(x, Q^2=10{\rm GeV}^2)=3.34x^{0.31}(1-x)^{5.06}(1-0.177x)~,
\label{eqn:typeB}\\
&&{\rm with}~~\Delta G(Q^2_{SMC})=5.64~,\nonumber
\end{eqnarray}
\item BBS model \cite{ref:Brod94}~:
\begin{eqnarray}
&&x\delta G(x, Q^2=4{\rm GeV}^2)=0.281\left\{(1-x)^4-(1-x)^6\right\}\nonumber\\
&&~~~~~~~~~~~~~~~~~~~~~~~~~~~~+1.1739\left\{(1-x)^5-(1-x)^7\right\}~,
\label{eqn:typeC}\\
&&{\rm with}~~\Delta G(Q^2_{SMC})=0.53~,\nonumber
\end{eqnarray}
\item no gluon polarization model~:
\begin{equation}
x\delta G(x, Q^2=10{\rm GeV}^2)=0~,~{\rm with}~~\Delta G(Q^2_{SMC})=0~,
\label{eqn:typeD}
\end{equation}
\end{enumerate}
where $Q^2_{SMC}$ is taken to be $10$GeV$^2$.
Among these distributions, $\Delta G$ of types (a) and (b) are large while
those of types (c) and (d) are small and zero, respectively.
The $x$--dependence of $x\delta G(x, Q^2)$ at $Q^2=10$GeV$^2$ and
$\delta G(x, Q^2)/G(x, Q^2)$ which are
evolved up to $Q^2=M_{\chi_2(c\bar c)}^2$ and $M_{\chi_2(b\bar b)}^2$ by
the Altarelli--Parisi equations, is depicted in Figures 2(A), (B) and (C),
respectively.
Most people\cite{ref:Altarelli89} who assume large $\Delta G$ take type (b) as
the $x\delta G$. As shown in Figure 2(A), the $x\delta G(x)$ of type (b)
has a peak at $x\approx 0.05$
and gradually decreases with increasing $x$ while that of (a) has a sharp
peak at $x\approx 0.04$ and rapidly decreases with $x$.
It is remarkable that the distribution of type (a) is consistent with
the experimental data of not only two--spin asymmetries
$A^{\pi^0}_{LL}(\stackrel{\scriptscriptstyle(-)}{p}\stackrel{}{p})$ for the
inclusive $\pi^0$--production\cite{ref:Morii94,ref:E581-1} but also recent
$A^{\gamma\gamma}_{LL}(pp)$ for the inclusive multi--$\gamma$ pair
production\cite{ref:E581-2} measured by the E581/704 Collaboration
at Fermilab by using longitudinally
polarized proton (antiproton) beams and longitudinally polarized proton
targets, even though it has large $\Delta G$ ($=5.32$) in a proton.
Here we examine above four types of polarzed gluon distribution functions to
discuss low--$p_T$ $\chi_J$--hadroproductions, though the type (b) has been
already ruled out by the data on two--spin asymmetries\cite{ref:Weber}.

\vspace{0.8cm}
\noindent
{\bf 2~~~Two--spin asymmetry $A_{LL}$}
\vspace{0.4cm}

Let us start by defining an interesting parameter called a two--spin asymmetry
for $\chi_J$--hadroproductions $A_{LL}^{\chi_J}$, which is given by
\begin{eqnarray}
A_{LL}^{\chi_J}~&=&~\frac{\left[d\sigma_{++}-d\sigma_{+-}+
d\sigma_{--}-d\sigma_{-+}\right]}
{\left[d\sigma_{++}+d\sigma_{+-}+
d\sigma_{--}+d\sigma_{-+}\right]}
\nonumber\\
&=&~\frac{d\Delta\sigma_{\chi_J}}{d\sigma_{\chi_J}}~.
\label{eqn:dfnALL}
\end{eqnarray}
Here $d\sigma_{+-}$, for instance, denotes that the helicity
of one beam particle is positive and the one of the other is negative.
Although several people have investigated $A_{LL}^{\chi_J}$
as a function of $\sqrt s$ \cite{ref:chi}, the experimental data are available
more easily as a function of the longitudinal fraction of produced particle
$x_L$. Then we investigate how the $x_L$--dependence of $A_{LL}^{\chi_J}$ for
$\chi_J$--hadroproductions in polarized
$pp$ collisions are affected by polarized gluon distribution functions.

Since the dominant source is the gluon--gluon fusion as
shown in Figure 1, the reaction is very sensitive to polarized gluon
distributions in a proton.
The $\chi_1$ state cannot be produced by this reaction due to
Yang's theorem. The spin--dependent and spin--independent subprocess cross
sections $\Delta\hat\sigma_{\chi_J}$ and $\hat\sigma_{\chi_J}$ for the
subprocess $G + G \to \chi_{0, 2}$ are straightforwardly calculated
as\cite{ref:Gastmans}
\begin{equation}
\Delta \hat\sigma_{\chi_0} = \frac{12\pi^2\alpha_S^2|R_1'|^2}{M_{\chi_0}^7}
\delta(1-\frac{M_{\chi_0}^2}{\hat s})~,
{}~~~~\hat\sigma_{\chi_0} = \frac{12\pi^2\alpha_S^2|R_1'|^2}{M_{\chi_0}^7}
\delta(1-\frac{M_{\chi_0}^2}{\hat s})
\label{eqn:direct1}
\end{equation}
for $\chi_0$--productions and
\begin{equation}
\Delta \hat\sigma_{\chi_2} = -\frac{16\pi^2\alpha_S^2|R_1'|^2}{M_{\chi_2}^7}
\delta(1-\frac{M_{\chi_2}^2}{\hat s})~,
{}~~~~\hat\sigma_{\chi_2} = \frac{16\pi^2\alpha_S^2|R_1'|^2}{M_{\chi_2}^7}
\delta(1-\frac{M_{\chi_2}^2}{\hat s})
\label{eqn:direct2}
\end{equation}
for $\chi_2$--productions. Here $R_1'$ is the derivative of the P--state wave
function at the origin. The spin--dependent and spin--independent differential
cross sections in terms of $x_L$ ($=2p_L/\sqrt s$) of produced $\chi_J$
($J=0, 2$) are
\begin{eqnarray}
\frac{d\Delta\sigma_{\chi_J}}{dx_L}&=&
\Delta C_{\chi_J}~\frac{\pi^2\alpha_S^2|R_1'|^2}{M_{\chi_J}^7}
\frac{\tau}{\sqrt{x_L^2+4\tau}}
\delta G(x_a, Q^2)~\delta G(x_b, Q^2)~,
\label{eqn:direct3}\\
\frac{d\sigma_{\chi_J}}{dx_L}&=&
C_{\chi_J}~\frac{\pi^2\alpha_S^2|R_1'|^2}{M_{\chi_J}^7}
\frac{\tau}{\sqrt{x_L^2+4\tau}}
G(x_a, Q^2)~G(x_b, Q^2)
\label{eqn:direct4}
\end{eqnarray}
with $x_a$ and $x_b$ being the momentum fraction in a proton and given as
\begin{equation}
x_a=\frac{x_L+\sqrt{x_L^2+4\tau}}{2}~,~~~
x_b=\frac{-x_L+\sqrt{x_L^2+4\tau}}{2}~,~~~
\tau\equiv \frac{M_{\chi_{0, 2}}^2}{s}~,
\label{eqn:direct5}
\end{equation}
where $\Delta C_{\chi_J}$ and $C_{\chi_J}$ are $12$ ($-16$) and $12$ ($16$) for
$\chi_0$ ($\chi_2$)--hadroproductions, respectively. By using these cross
section formulas, the explicit form of two--spin asymmetries
$A_{LL}^{\chi_{0 (2)}}(pp)$ can be written as
\begin{equation}
A_{LL}^{\chi_{0 (2)}}(pp)=(-)~
\frac{\delta G(\frac{x_L+\sqrt{x_L^2+4\tau}}{2}, Q^2)}
{G(\frac{x_L+\sqrt{x_L^2+4\tau}}{2}, Q^2)}
{G(\frac{-x_L+\sqrt{x_L^2+4\tau}}{2}, Q^2)}~.
\label{eqn:direct6}
\end{equation}
It is remarkable that at low--$p_T$ regions, $A_{LL}^{\chi_{0 (2)}}(pp)$ which
is originated mainly from gluon--gluon fusion of Figure 1, is directly
proportional to polarized gluon distributions $\delta G(x)$. Therefore,
$A_{LL}^{\chi_{0 (2)}}(pp)$ sensetively depends on $\delta G(x)$. Namely,
$A_{LL}^{\chi_{0 (2)}}(pp)$ at low--$p_T$ regions is a very good physical
parameter to test the behavior of $\delta G(x)$.
To derive the differential cross sections in
Eqs.(\ref{eqn:direct3}) and (\ref{eqn:direct4}),
the higher order QCD corrections, the quarkonium binding effect and the
corrections due to relativistic effects of the quarkonium should be taken into
account. However, these uncertainties are generally included in
{\it so--called} ``K--factor",
and they would cancel if those corrections contribute to the denominator and
numerator of Eq.(\ref{eqn:direct6}) alike.

As the PHENIX detector which is to work at RHIC is dedicated to have a special
device for detecting leptons and $\gamma$'s\cite{ref:RHIC}, the observable
signal for $\chi_{0, 2}$--productions is expectded to be two--muon decays of
$^3S_1$ coming from radiative decays $\chi_{0, 2}\to {^3S}_1+\gamma$.
Since the differential cross sections $d\Delta\sigma_{\chi_2}/dx_L$
and $d\sigma_{\chi_2}/dx_L$ are almost equal to
$d\Delta\sigma_{\chi_0}/dx_L$ and $d\sigma_{\chi_0}/dx_L$,
respectively, the $\chi_2$ is easier to be detected because it has a
larger branching ratio for the radiative decay than that of the $\chi_0$.

Using the polarized gluon
distributions of four types given in Eqs.(\ref{eqn:typeA})--(\ref{eqn:typeD})
and taking $Q^2$ as $M_{\chi_2}^2$,
we estimate the $A_{LL}^{\chi_2}(pp)$ as a function of $x_L$ for
$\chi_2$--productions of not only the charmonium $\chi(c\bar c)$ but also
the bottomonium $\chi(b\bar b)$ states
at relevant RHIC energies, $\sqrt s=50$, $200$ and $500$GeV.
Although the cross section of $\chi(b\bar b)$ is smaller than that of
$\chi(c\bar c)$ about by 10$^{-3}$, $\chi(b\bar b)$ has different values of
$x_a$ and $x_b$ given in Eq.(\ref{eqn:direct5}) from those of $\chi(c\bar c)$
productions, and so we can obtain not only
$Q^2$--dependence but also $x$--dependence of the polarized gluon
distributions by examining $\chi(b\bar b)$ states together with
$\chi(c\bar c)$ states at the same time.
We show the results of $A_{LL}^{\chi_2}$
calculated  for (A) $c\bar c$ and (B) $b\bar b$ states at $\sqrt s=50$,
$200$ and $500$GeV in Figures 3, 4 and 5, respectively.

In the case of $\chi_2(c\bar c)$, there
is a tendency that the smaller $\sqrt s$ (namely $\tau$) is,
the larger $A_{LL}^{\chi_2(c\bar c)}$ become. In particular, for
$\sqrt s=50$GeV, it might be easy to examine not only the magnitude
$\Delta G$ but also the behavior $\delta G(x)$ of the gluon polarization.
In addition, $A_{LL}^{\chi_2(c\bar c)}(pp)$ at $\sqrt s=200$ and $500$GeV is of
use to know information on polarized gluon distributions, provided
experiments are carried out with the high precision.

As for $\chi_2(b\bar b)$--productions, the dependence of
$A_{LL}^{\chi_2(b\bar b)}(pp)$ on $\sqrt s$ has a similar tendency as that of
$A_{LL}^{\chi_2(c\bar c)}(pp)$.
However, the behavior of $A_{LL}^{\chi_2(b\bar b)}(pp)$ calculated at
$\sqrt s=50$GeV by using our $\delta G(x)$ (type (a)) differs from that at the
other $\sqrt s$. This is because for small $x_L$ ($\lsim 0.3$), $x_a$ and $x_b$
($=x_a-x_L$) are taken to be about $0.2-0.4$ and hence
$\delta G(x_a)/G(x_a)$ and $\delta G(x_b)/G(x_b)$ have around the
minimal value as shown in Figure 2(C). Accordingly, Figure 3(B) indicates that
it is rather difficult to distinguish our large $\delta G(x)$ (type (a)) from
no polarization $\delta G(x)=0$ (type (d)) in the region $x_L\lsim 0.3$.
At other $\sqrt s$, \ie $\sqrt s=200$ and $500$GeV, since the difference of
$A_{LL}^{\chi_2(b\bar b)}$ for any type of polarized gluon
distributions is relatively larger than that of $A_{LL}^{\chi_2(c\bar c)}$,
it might be useful to examine the magnitude and the $x$--dependence of
$\delta G$, though the differential cross section of $\chi_2(b\bar b)$ is
smaller than that of $\chi_2(c\bar c)$.
Both $A_{LL}^{\chi_2(c\bar c)}$ and $A_{LL}^{\chi_2(b\bar b)}$ for the no gluon
polarization (type (d)) are almost zero in Figures (4), (5) and (6).

We find that the low--$p_T$ $\chi_2$--productions allow to give the clear test
as a probe of the magnitude and the $x$--dependence of the gluon polarization
because $A_{LL}^{\chi_2}$ for those reactions is directly proportional to
$\delta G(x)/G(x)$.
We have not taken up here the large--$p_T$ $\chi_J$--hadroproductions though
they are also interesting for getting the informations on polarized
gluon\cite{ref:Morii95}.

\vspace{0.8cm}
\noindent
{\bf 3~~~Summary}
\vspace{0.4cm}

We have calculate the two--spin asymmetries for the
low--$p_T$ productions for $\chi_0$ and $\chi_2$ of the $c\bar c$ state and the
$b\bar b$ state.
The experimental data on charmonium and bottomonium $\chi_J$--productions in
polarized $pp$
collisions at RHIC will be very helpful to put constraints on the magnitude
and the behavior of the gluon polarization, and improve our understanding of
the proton structure.
In particular, the low--$p_T$
productions for $\chi_2$ can give a very useful information on polarized gluon
distributions because the two--spin asymmetry for these reactions is simply
proportional to $\delta G(x)/G(x)$.
We hope that our predictions will be tested in forthcoming
RHIC spin experiments.

\vspace{1cm}
\noindent
{\bf Acknowledgments}
\vspace{0.5cm}

The authors are grateful to Profs. A. Masaike and S. Nurushev
for useful comments on experimental issues at RHIC.
We would like to acknowledge fruitful discussions on the subject of this work
with Profs. G. Ramsey and D. Sivers.

\vfill\eject

\begin{center}
{\large \bf Figure captions}
\end{center}
\begin{description}
\item[Fig. 1:] A diagrams of the dominant process for low--$p_T$ productions
of $p~p\rightarrow \chi_J~X$ at lowest order QCD.

\vspace{2em}

\item[Fig. 2:] The $x$--dependence of (A) $x\delta G(x, Q^2)$ at
$Q^2=10$GeV$^2$, (B) $\delta G(x, Q^2)/G(x, Q^2)$ at
$Q^2=M_{\chi_2(c\bar c)}^2$ and (C) $\delta G(x, Q^2)/G(x, Q^2)$ at
$Q^2=M_{\chi_2(b\bar b)}^2$ for various types (a)--(d) given by
Eqs.(\ref{eqn:typeA})--(\ref{eqn:typeD}).

\vspace{2em}

\item[Fig. 3:] Two--spin asymmetries (A) $A_{LL}^{\chi_2(c\bar c)}(pp)$
and (B) $A_{LL}^{\chi_2(b\bar b)}(pp)$ for $\sqrt s=50$GeV,
calculated with various types of $\delta G(x)$, as a function of longitudinal
momentum fraction $x_L$ of $\chi_2(c\bar c)$ and $\chi_2(b\bar b)$.
The solid, dashed, small--dashed and dash--dotted lines indicate the results
using types (a), (b), (c) and (d) in Eqs.(\ref{eqn:typeA}), (\ref{eqn:typeB}),
(\ref{eqn:typeC}) and (\ref{eqn:typeD}), respectively.

\vspace{2em}

\item[Fig. 4:] Two--spin asymmetries (A) $A_{LL}^{\chi_2(c\bar c)}(pp)$ and
(B) $A_{LL}^{\chi_2(b\bar b)}(pp)$ for $\sqrt s=200$GeV.
Various lines represent the same as in Figure 3.

\vspace{2em}

\item[Fig. 5:] Two--spin asymmetries (A) $A_{LL}^{\chi_2(c\bar c)}(pp)$ and
(B) $A_{LL}^{\chi_2(b\bar b)}(pp)$ for $\sqrt s=500$GeV.
Various lines represent the same as in Figure 3.
\end{description}

\begin{thebibliography}{99}
\bibitem{ref:RHIC}
RSC proposal, BNL, R 5, August (1992).
\bibitem{ref:EMC88}
J. Ashman \etal, \Journal{\PLB}{206}{364}{1988};
\Journal{\NPB}{328}{1}{1989}.
\bibitem{ref:SMC94}
D. Adams \etal, \Journal{\PLB}{329}{399}{1994}.
\bibitem{ref:Ellis74}
J. Ellis and R. L. Jaffe, \Journal{\PRD}{9}{1444}{1974};
\Journal{\PRD}{10}{1669}{1974}.
\bibitem{ref:anomaly}
G. Altarelli and G. G. Ross, \Journal{\PLB}{212}{391}{1988};
R. D. Carlitz, J. C. Collins and A. H. Mueller, \Journal{\PLB}{214}{229}{1988};
A. V. Efremov and O. V. Teryaev, in {\it Proceedings of the International
Hadron Symposium}, ed. Fischer \etal,
(Czechoslovakian Academy of Scinece, Prague, 1989).
\bibitem{ref:Baier83}
R. Baier and R. R\"uckl, \Journal{\ZPC}{19}{251}{1983};
E. W. N. Glover, A. D. Martin and W. J. Stirling,
\Journal{\ZPC}{38}{473}{1988}.
\bibitem{ref:Altarelli89}
G.Altarelli and W.J.Stirling, {\it Part. World} {\bf 1} 40, (1989);
Z. Kunszt, \Journal{\PLB}{218}{243}{1989};
J. Ellis, M. Karliner and C. T. Sachrajda, CERN--TH--5471/89;
M. Gl\"uck, E. Reya and W. Vogelsang, \Journal{\PRD}{45}{2552}{1992}.
\bibitem{ref:Cheng90}
H. Y. Cheng and S. N. Lai, \Journal{\PRD}{41}{91}{1990}.
\bibitem{ref:Rams91}
G. Ramsey and D. Sivers, \Journal{\PRD}{43}{2861}{1991}.
\bibitem{ref:Morii94}
T. Morii, S. Tanaka and T. Yamanishi, preprint KOBE--FHD--94--08 (1994), to be
published in {\it Z. Phys.} {\bf C}.
\bibitem{ref:Brod94}
S. J. Brodsky, M. Burkardt and I. Schmidt, preprint 6087 (1994), to be
published in {\it Nucl. Phys.} {\bf B}.
\bibitem{ref:Gehrmann}
T. Gehrmann and W. J. Stirling, Durham preprint DTP/94/38 (1994), to be
published in {\it Z. Phys.} {\bf C}.
\bibitem{ref:E581-1}
D. L. Adams \etal, \Journal{\PLB}{261}{197}{1991}.
\bibitem{ref:E581-2}
D. L. Adams \etal, \Journal{\PLB}{336}{269}{1994}.
\bibitem{ref:Weber}
W. Vogelsang and A. Weber, \Journal{\PRD}{45}{4069}{1992}.
T. Yamanishi, Doctoral thesis, Kobe University, March (1994).
\bibitem{ref:chi}
J. L. Cortes and B. Pire, \Journal{\PRD}{38}{3586}{1988};
M. A. Doncheski and R. W. Robinett, \Journal{\PLB}{248}{188}{1990}.
\bibitem{ref:Gastmans}
R. Gastmans and T. T. Wu, {\it The Ubiquitous Photon Helicity Method for
QED and QCD} (Oxford, 1990).
\bibitem{ref:Morii95}
T. Morii, S. Tanaka and T. Yamanishi, preprint KOBE--FHD--95--02 (1995).
\end{thebibliography}
\end{document}